\documentclass[prl,twocolumn,superscriptaddress,showpacs,
twoside,floatfix]{revtex4}
\usepackage{dcolumn}

\usepackage{epsfig}
\newcommand {\la} {\langle}
\newcommand {\ra} {\rangle}
\newcommand {\beq} {\begin{eqnarray}}
\newcommand {\eeqn} [1] {\label{#1} \end{eqnarray}}%
\newcommand {\eol} {\nonumber \\}
\newcommand {\ve} [1] {\mbox{\boldmath $#1$}}

\begin{document}
%
%

\title{
A new insight into the observation 
of  spectroscopic strength  reduction in atomic nuclei:
implication  for the physical 
meaning of spectroscopic factors
 }

\author{
N.\ K.\ Timofeyuk
}                 

\affiliation{
 Department of Physics, 
University of Surrey, Guildford,
Surrey GU2 7XH,  UK
}

\date{\today}


\begin{abstract}

Experimental studies of one nucleon knockout from magic nuclei
suggest that their nucleon orbits are not fully occupied. This
conflicts a commonly accepted view of the shell closure associated
with  such nuclei.
The conflict can be reconciled if the overlap   
between initial and final nuclear states in a knockout reaction
are calculated by a non-standard method. 
The method employs an inhomogeneous equation based on
correlation-dependent effective nucleon-nucleon (NN)  interactions and allows
the simplest wave functions, in which all nucleons occupy only
the lowest nuclear orbits, to be used.
The method also reproduces the  recently established relation
between reduction of spectroscopic strength,
observed in knockout reactions on other nuclei, and nucleon
binding energies. The implication  of the inhomogeneous equation
method  
for the physical meaning of spectroscopic factors is discussed.


\end{abstract}

\maketitle

The concept of "magic numbers" of neutrons and protons making up a nucleus 
is fundamental to our understanding of a wide range of phenomena from 
the properties and binding energies of nuclei themselves to the relative 
abundance of  elements in the universe \cite{Mey}. The observed "magic numbers" 
are usually explained by a model (the Shell Model) in which independent 
nucleons fill single particle energy levels in a mean field according 
to the Pauli exclusion principle. Such a picture is similar to
electronic structure of atoms responsible for organising the chemical elements
into the Periodic Table.
It has been found, however, that the cross sections of
the $(e,e'p)$ reactions on  the closed shell  nuclei   $^{16}$O,
$^{40,48}$Ca, $^{208}$Pb 
are 50-60 $\%$ smaller than those expected
from the independent particle model \cite{Pan97}.
Direct reaction theories of the $(e,e'p)$ reaction predict that its 
cross section  depends on the Spectroscopic Factor (SF) which is
a measure of  the occupancy of single proton levels 
in the target nucleus.  The observed reduction of SFs  
appears to contradict the traditional view of
$^{16}$O,
$^{40,48}$Ca, $^{208}$Pb as doubly magic  nuclei.


Away from the closed shells, nuclei 
such  as  $^{7}$Li, $^{12}$C,  $^{30}$Si, $^{31}$P, $^{51}$V and
$^{90}$Zr  also reveal a similar reduction of SFs as compared to prediction of
the 0$\hbar\omega$ shell model   \cite{Kra01}.
The SF  reduction is also found for other nuclei
in a recent analysis of the
$(d,p)$ and $(p,d)$ reactions \cite{Lee06}, in  which 
the bound state wave functions of the transferred neutron are  fixed by
modern Hartree-Fock calculations and have   shapes  similar
to those derived from $(e,e'p)$.

Recently,   SF  studies with radioactive beams
have revealed a new phenomenon.
It turned out that   reduction of experimental SFs $S_{exp}$,
detemined as ratios of the measured to theoretical cross sections,
from the theoretical values
$S_{th}$, obtained in the shell model, 
depends on the separation energy of the removed nucleon  and on the 
nucleon type. It has been also discovered that
the SF reduction factor $R_s=S_{exp}/S_{th}$  is concentrated
around a straight line when plotted as a function of  
the  difference 
between proton ($S_p$) and neutron ($S_n$) separation energies, 
$\Delta S$, taken
as $S_p-S_n$    and $S_n-S_p$  for proton and  neutron knockout,
respectively \cite{Gad08}.

It is known that 
$S_{th}$  agrees better with  $S_{exp}$
if the model space, in which $S_{th}$ is calculated, is increased, 
or in other words,
if particle-hole excitations are allowed.
Thus, a six-shell treatment of $^{16}$O shows that the percentage of the
0$\hbar\omega$ component in it is $\sim$48-60$\%$ \cite{War92} and that
the $^{16}$O SF changes from the 0$\hbar \omega$
value of 2 to 1.65 when the model space  increases to 4$\hbar \omega$
\cite{Bro02}.
However, it is still higher than the $(e,e'p)$  value of 1.27(13) \cite{Kra01}
suggesting that more major shells should be added  to
the   model space, which   contradicts the   view of $^{16}$O
as a double magic nucleus. 
The contributions from missing  model spaces can be recovered by using
correlated wave functions in {\it ab-initio}   approaches. Indeed,
the $^{7}$Li proton SF 
calculated in the Variational Monte Carlo (VMC)  
agrees very well
 with $S_{exp}$ from $(e,e'p)$ \cite{Lap99}. 
However, for $^{8,9}$Li, $^8$B and $^9$C
the  SF  reduction obtained by VMC calculations
is not sufficient (see Table I).
Also, the {\it ab-initio} calculations are  feasible only for light nuclei 
while the SF reduction
is observed for  nuclei as heavy as $^{208}$Pb.

In this letter, I  show 
that  it is possible to reconcile the double magic
nature of $^{16}$O  with the observed
60$\%$ reduction of its spectroscopic strength 
and at the same time to explain 
the observed $R_s(\Delta S)$ dependence if an alternative method to
calculating SFs is used. This method allows minimal shell model spaces
to be used
and accounts  automatically for excluded orbits.
It can be applied to any 
nucleus and can be  introduced into  
existing shell model codes including those used by the   
community of nuclear experimentalists studying one nucleon removal reactions.
Below, I present this method,  emphasize its importance 
for explaining the  phenomenon  of SF reduction
and  present  numerical results for   $A < 16$ nuclei.

The theoretical SF for one-nucleon removal, $S_{lj}$, 
is defined in a model independent way as
the norm of the radial overlap function $I_{lj}(r)$ with orbital momentum $l$
and angular momentum $j$, calculated  between the
  wave functions $\Psi_{J_{B}}$ and $\Psi_{J_A}$
of two neighbouring 
nuclei  $B = A-1$ and $A$ with the total spin $J_B$ and $J_A$: 
\beq
 I_{lj}^{DE}(r)  =
A^{\frac{1}{2}}
\la[[ 
Y_{l}(\hat{\ve{r}}) \otimes \chi_{\frac{1}{2}} ]_j
\otimes \Psi_{J_B}]_{J_A}|\Psi_{J_A}\ra. 
\eeqn{vff}
All available shell model codes calculate $S_{lj}$ from  $I_{lj}(r)$
obtained by direct evaluation (DE) of Eq. (\ref{vff}) 
using some  model wave functions in truncated model spaces.
The input to these shell model calculations includes matrix elements
of the effective nucleon-nucleon (NN) interaction fitted to a range of nuclear
spectra.  
They carry  no information about  the radial shapes of 
$I_{lj}(r)$,  crucial for calculating    
one nucleon removal cross sections.  In most applications, these shapes are   
found from  the separation energy prescription, not related
to the shell model NN matrix elements.

An  alternative method  to calculate $I_{lj}(r)$  is 
to solve the inhomogeneous equation (IE) 
\beq
\la\Psi_B|\hat{T}_A-\hat{T}_B-E_A+E_B|\Psi_A \ra 
= \la\Psi_B|V_B-V_A|\Psi_A  \ra,
\eeqn{IE}
 originally introduced by 
Pinkston and Satchler
\cite{Pin65}.
Here
$\hat{T}_i$ and $V_i$ are the kinetic and potential energy operators 
while $E_i$ is the total energy of nucleus $i$. The r.h.s. of (\ref{IE}) is
treated as known. 
Eq. (\ref{IE})  generates $I_{lj}(r)$
which automatically have the correct asymptotic shape, a feature crucial for
transfer reactions but not so for binding energy calculations. Earlier
explorations of this method, reviewed in \cite{Sat83}, were based
on separating the mean field part out of $V_{i }$ 
and keeping only the valence nucleon space. They gave little information of
utility
of the method and were abandoned  before 1980s.
Later, a different strategy,   applied in \cite{Muk90,Tim}
to calculating the source
term $\la\Psi_B|V_B-V_A|\Psi_A  \ra$,  resulted
in SFs different from traditional shell model values.
Neither the Pinkston-Satchler approach nor that of
Refs. \cite{Muk90,Tim} have been considered in the context
of the SF reduction phenomenon as both were used at the time when
the $R_s(\Delta S)$ dependence was not known. Here, I prove the legitimacy of
the method of Ref. \cite{Tim}  and  show its
relevance to the SF reduction.

According to \cite{Tim}, the solution of Eq. (\ref{IE}) is
\beq
I_{lj}^{IE}(r)  =
A^{\frac{1}{2}}
\la[[ \frac{G_l(r,r')}{rr'}
Y_{l}(\hat{\ve{r}'}) \otimes \chi_{\frac{1}{2}} ]_j
\otimes \Psi_{J_B}]_{J_A}||\hat{{\cal V}}||\Psi_{J_A}\ra, 
\eol
\eeqn{ineq}
where integration over $r'$ is implied and
$G_l(r,r')$ is the  Green function for a bound nucleon in the 
field of a point charge $Z_B$ corresponding
to the momentum $i\kappa$,
\beq
G_l(r,r')=-\frac{2\mu}{\hbar^2 \kappa} e^{-\frac{\pi i(l+1+\eta)}{2}}
F_l(i\kappa r_<) W_{-\eta,l+\frac{1}{2}}(2\kappa r_>). \eol
\eeqn{GF}
Here 
 $\kappa=(2\mu\epsilon/\hbar^2)^{1/2}$,
$\epsilon = E_B - E_A$, 
  $\mu$ is reduced mass,
$\eta = Z_BZ_N e^2 \mu/\hbar^2 \kappa$, $F$ is the regular Coulomb
function and $W$ is the Whittaker function. Also,
$\hat{{\cal V}} = V_A - V_B - Z_BZ_Ne^2/r$ and $V_x = \sum_{i<j}^x v_{ij}$.
In both Eqs. (\ref{vff})
and (\ref{ineq}), $r$ ($r'$) 
is the distance between the centre-of-mass of $B$ and 
the removed nucleon, and $Y_l$ is the spherical function.
The advantage of (\ref{ineq}) is that it 
guarantees the
correct asymptotic form for $I^{IE}_{lj}(r)$ when the experimental value of 
$\epsilon$
is used, whatever $\Psi_{J_{B}}$ and $\Psi_{J_A}$ are.

Eq. (\ref{ineq}) was obtained assuming that $\Psi_{J_{B}}$ and $\Psi_{J_A}$
are exact solutions 
of the many-body 
Schr\"odinger equation and that $\hat{{\cal V}}$
contains bare realistic NN interactions. In this case,
$I_{lj}^{DE}(r)$ and
$I_{lj}^{IE}(r)$, and the corresponding SFs $S_{lj}^{DE}$ and
$S_{lj}^{IE}$, 
should be equal. 
However, usually $\Psi_{J_{B}}$ and $\Psi_{J_A}$ are replaced by
model   wave functions in truncated spaces.
This raises the question about what should
be used for $\hat{{\cal V}}$. To answer it,  I consider
an exact nuclear wave function $\Psi$   
constructed from an uncorrelated state $\Phi$, defined in some
truncated model space, for example,
using the Unitary Correlation Operator Method   \cite{Fel98}:
\beq
|\Psi \ra =  C |\Phi \ra = \exp \{-i \sum_{i<j}^A g_{ij}\} |\Phi \ra.
\eeqn{wfucom}
Here $C$ is the unitary correlator designed to shift nucleons away from 
each other whenever their uncorrelated positions are within the repulsive
NN core. $\Phi$ is found from an effective Hamiltonian  that contain 
effective interactions $V^{\rm eff}$
consisting of $\hat {V}= C^{\dagger}VC$ and the terms
arising from the kinetic energy operator \cite{Fel98}.
If  wave functions from Eq. (\ref{wfucom})
are used in Eq. (\ref{ineq}), then
\beq
\la \Psi_B | \hat{ {\cal V}}|\Psi_A \ra  
= \la \Phi_B | C_B^{\dagger}(V_A - V_B)C_A|\Phi_A \ra  
\eol
= \la \Phi_B |   V_N  C_{NB}|\Phi_A \ra= 
\la \Phi_B | \tilde {V}^{\rm  eff}|\Phi_A \ra,
\eeqn{abucom}
where $C_A = C_B C_{NB}$, $C_{NB} = \exp\{-i \sum_{i=1}^{B} g_{iN}\}$
and $V_{N } = \sum_{i=1}^{B}v_{iA}$, 
assuming for simplicity that Coulomb interaction is absent. 
Eq. (\ref{abucom}) tells us that the effective interaction 
$\tilde {V}^{\rm  eff}$ that approximates $\hat{ {\cal V}}$  
when modelling   $I_{lj}^{IE}(r)$ using uncorrelated model
functions $\Phi_B$ and $\Phi_A$, differs from the effective interaction 
$V^{\rm eff}$  that generates them.
Moreover, $\Phi_B$ and $\Phi_A$ depend only on matrix elements
$\la \psi_{\alpha_1}(\ve{r}_1)\psi_{\alpha_2}(\ve{r}_2)  | 
v^{\rm eff}(r_{12})| \psi_{\alpha_3}(\ve{r}_1) \psi_{\alpha_4}(\ve{r}_2) \ra$
in a chosen truncated space, where $\psi_{\alpha}(\ve{r})$
is a single-particle wave function in the state $\alpha$. Hence, 
$I_{lj}^{IE}(r)$ and $S_{lj}^{IE}$ depend on them as well. But in 
addition, they also depend on  matrix elements of $\tilde {V}^{\rm  eff}$,
$\la G_l(r,r_1)/(r_1r) \psi_{\alpha_2}(\ve{r}_2)  | 
\tilde{v}^{\rm eff}(r_{12})| \psi_{\alpha_3}(\ve{r}_1) 
\psi_{\alpha_4}(\ve{r}_2) \ra$
(if centre-of-mass motion is neglected), that carry information
about coupling to missing model spaces. This conclusion follows
from the Green function expansion onto 
complete set $\{\psi_{\alpha}(\ve{r})\}$, which includes
states from both truncated and missing spaces. Thus, these matrix elements
are not constrained by binding energy calculations 
and  must be constrained by some other means.


A  quantity that can serve as a reference to callibrate 
$\tilde{V}^{\rm eff}$ is the Asympotic Normalization Coefficient (ANC).
It  
determines the magnitude of the $I_{lj}(r)$ tail \cite{BBD77},
depends on the same operator $\tilde{V}^{\rm eff}$ and  can be  
determined from peripheral transfer experiments. 
In \cite{MT90}, the vertex constants, related to the ANCs by a trivial
relation \cite{BBD77}, were studied for $0p$-shell nuclei in
the oscillator $0\hbar\omega$ shell model.
It was   found 
that  reasonable agreement between measured
and calculated vertex constants  can be achieved if 
a version of the M3Y potential,
constructed in \cite{M3Y}
to fit the oscillator matrix elements derived from the NN scattering
phase shifts, is used for $\tilde{V}^{\rm eff}$. Below, I use this
interaction (labled as M3YE) to calculate  $S_{lj}^{IE}$. I show
that they are reduced with respect to $S_{lj}^{DE}$ and, at the same time,
 are closer to experimental SFs.

\begin{figure}[t]
\centerline{\hspace{-0.3 cm}
\psfig{figure=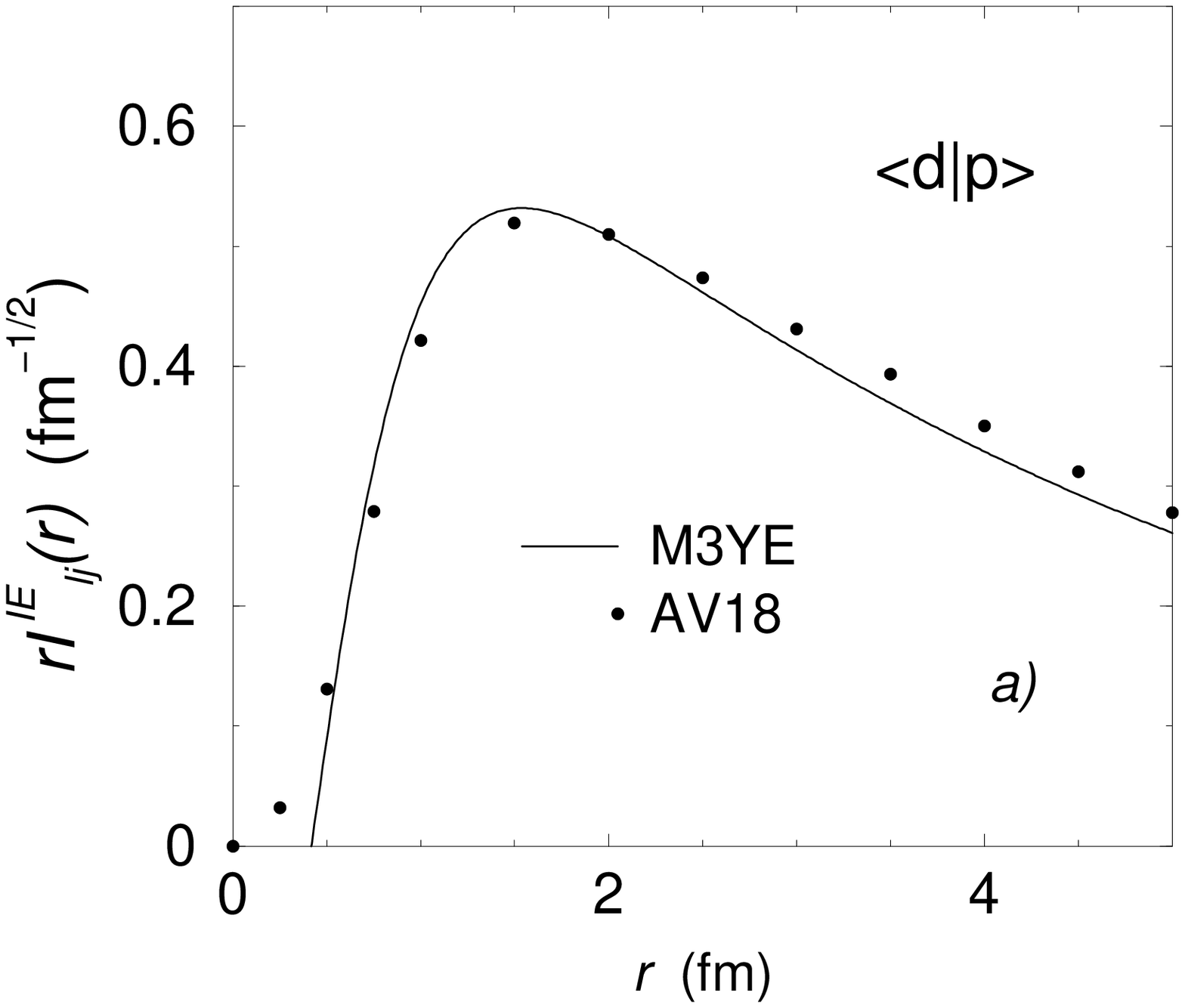,width=0.23\textwidth}
        \hspace{-0.0 cm}
        \psfig{figure=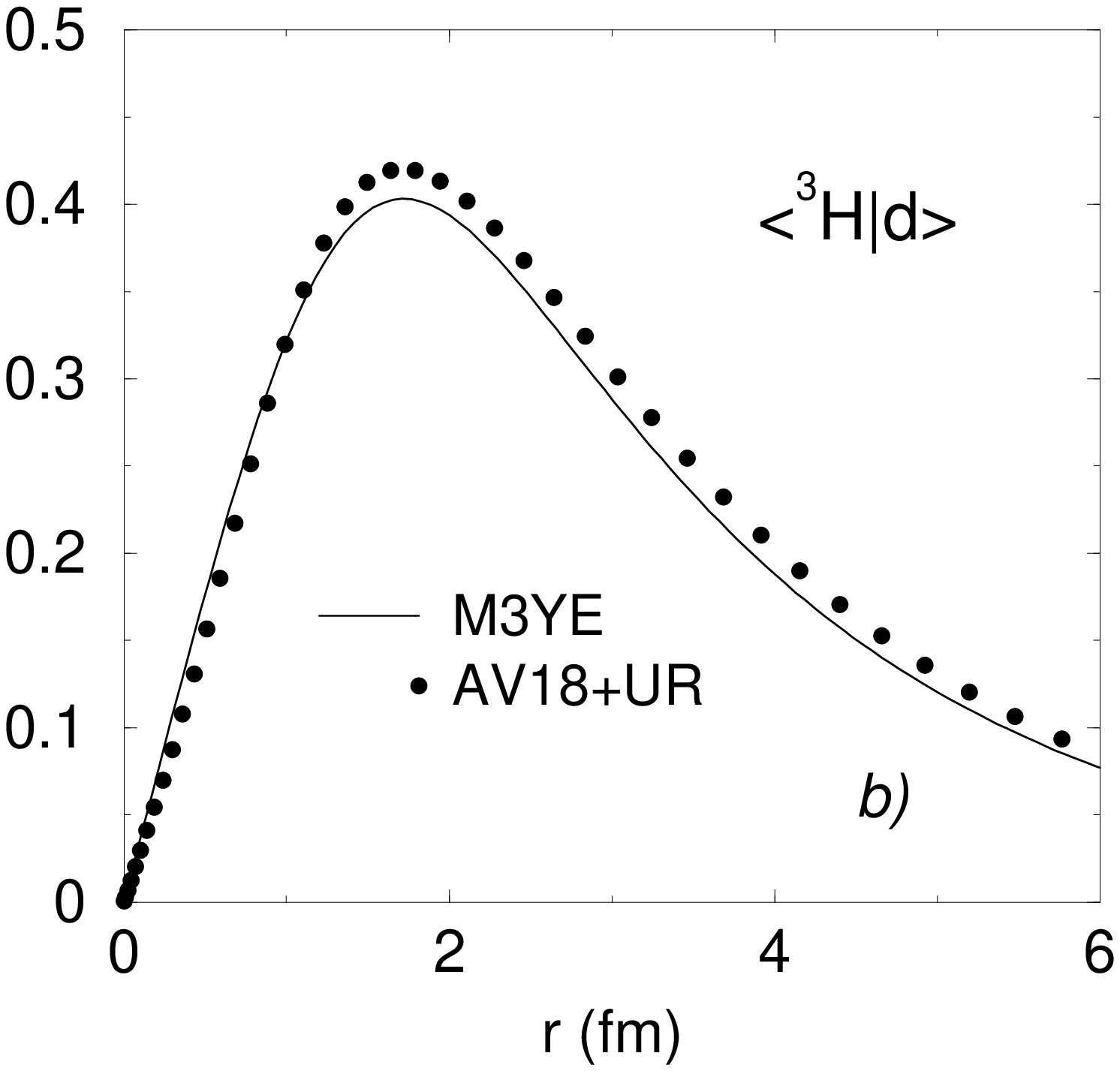,width=0.215\textwidth}
        }
        \centerline{
        \hspace{-0.3 cm}
        \psfig{figure=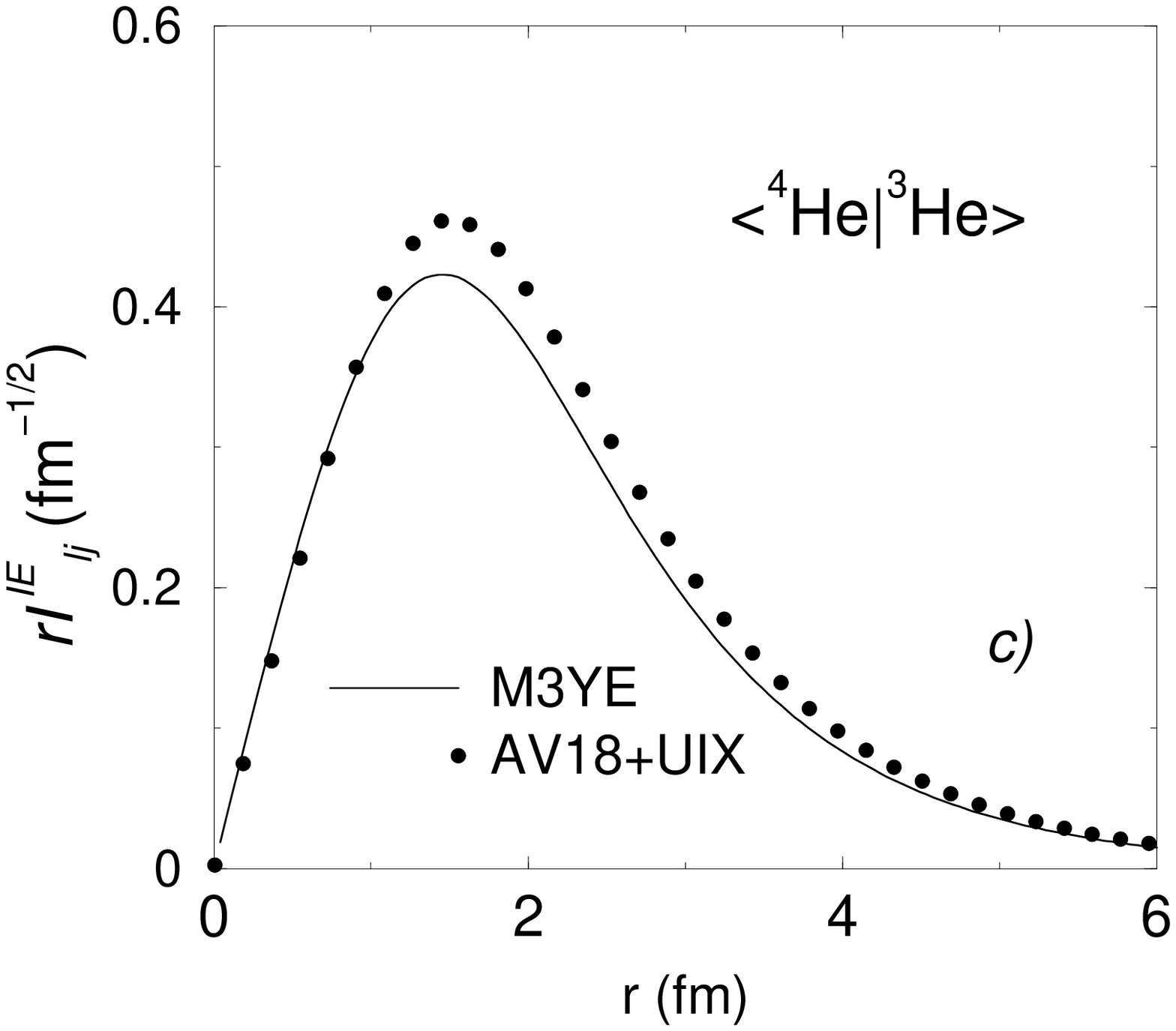,width=0.23\textwidth}
        \hspace{-0.0 cm}
        \psfig{figure=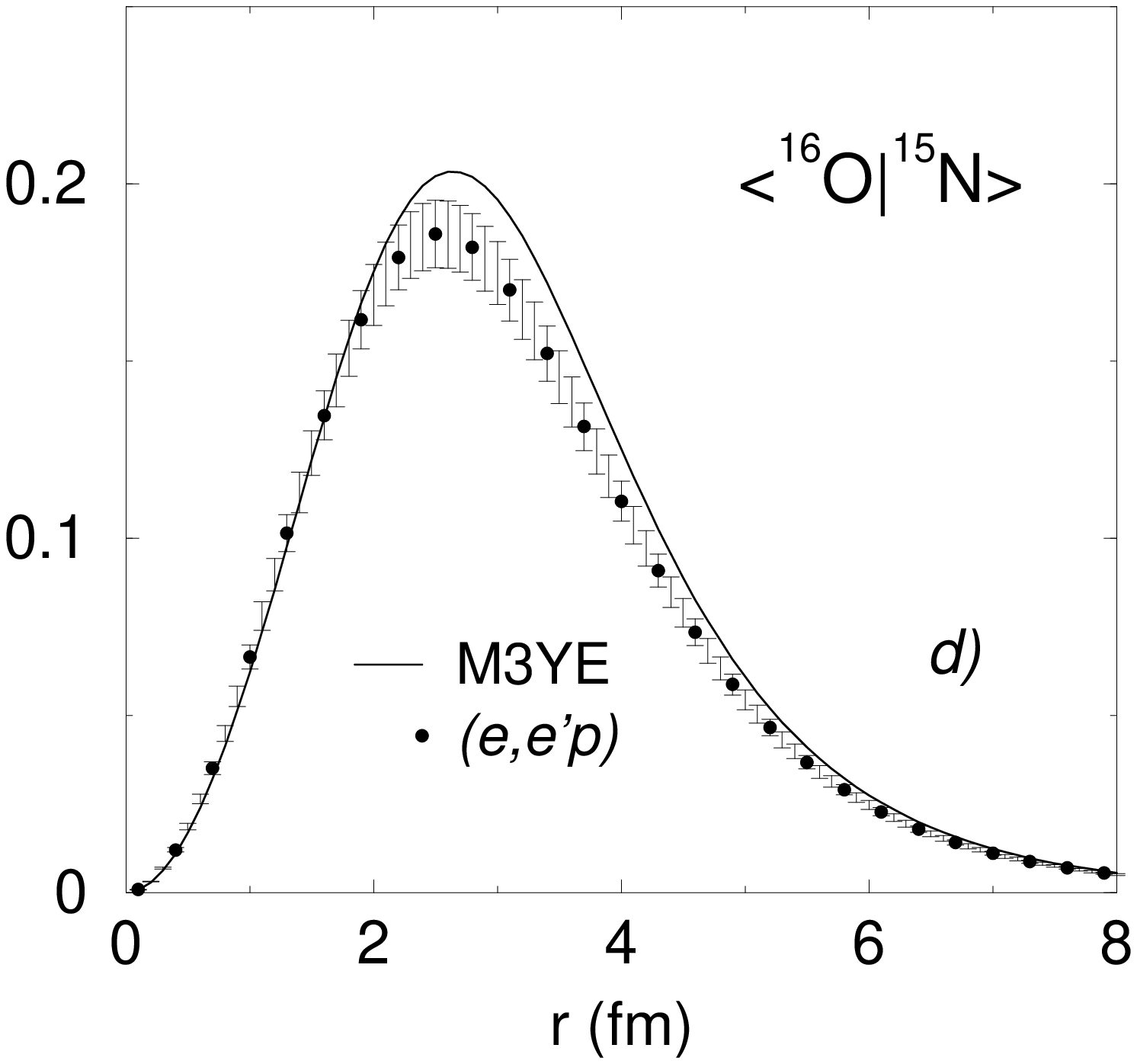,width=0.215\textwidth}
        }
\caption{The overlap functions $rI^{IE}(r)$ calculated for M3YE 
 for 
 (a) $\la d|p\ra$, (b) $\la^3$H$|d\ra$, (c)   $\la^4$He$|^3$He$\ra$ and 
 (d) $\la^{16}$O$|^{15}$N$\ra$ in comparison to (a) 
  the realistic
deuteron wave function obtained with AV18 potential in \cite{AV18},
(b,c)    {\it ab-initio} overlaps obtained with 
AV18+UR( or UIX) interaction in \cite{Kie97,Viv05} and (d) the overlap
function derived from $(e,e'p)$ reaction \cite{Kra01}.}
\end{figure}

First of all, I test the method as applied to the well understood 
$A$=2 system,  for which $I_{lj}^{IE}(r)$  is
the deuteron  wave function   and  satisfies
\beq
rI_{0s}^{IE}(r)=\int_0^{\infty} dr' \, r' G_0(r,r') 
\tilde{V}^{\rm eff}(r') \varphi_{0s}(r'),
\eeqn{dpn}
where $\varphi_{0s}$ is 
the $0s$ oscillator wave function. $rI_{0s}^{IE}(r)$,  calculated 
with  M3YE  for $\tilde{V}^{\rm eff}$ and with  oscillator
radius  $r_{osc}$=1.51 fm, 
 is close to the realistic deuteron
wave function generated by the NN potential AV18 \cite{AV18} (see Fig. 1a).
Its norm,
$S^{IE}$=0.91,
is    close  to the $s$-wave probability of 0.94 established
in the deuteron.

For closed shell nuclei, $I_{lj}^{IE}(r)$ depends only
on
 $\tilde{V}^{\rm eff}$ and  does not depend on the effective
interactions 
 determining their  energies. Thus, 
the SFs for these nuclei, together with their ANCs, can serve in the future
as a reference for callibrating
the interaction $\tilde{V}^{\rm eff}$. Here, I calculate
the overlaps $\la ^3$H$|d\ra$, $\la ^4$He$|^3$He$\ra$ and
$\la^{16}$O$|^{15}$N$\ra$, involving closed shell nuclei, 
using  M3YE. 
Only one
Slater determinant has been used
for $\Phi_A$ and $\Phi_B$, which are divided by the 
$0s$ centre-of-mass motion wave 
function. The
$r_{osc}$ is
chosen to be
1.53 for $^3$H and $^3$He, 1.33 fm for $^4$He and 1.8 fm for
$^{15}$N and $^{16}$O to reproduce their r.m.s. radii.
For $A$=3 and $A$=4 $I_{lj}^{IE}(r)$  are 
slightly smaller than the
{\it ab-initio} overlaps from \cite{Kie97,Viv05}    (see
Fig.1 b-d) but for $A$=16 
$I_{lj}^{IE}(r)$ is slightly larger than the overlap function derived from the
$(e,e'p)$ knockout \cite{Kra01}. In both cases, $S^{IE}$  are reduced
with respect to $S^{DE}$ (see Table I).

\begin{table}
\caption{
 $S^{IE} = S^{IE}_{p1/2} +S^{IE}_{p3/2} $  
 calculated 
with M3YE and RM3YE  in comparison to
$S^{DE}$, experimental values $S_{exp}$ 
\cite{Kra01,End03,Lap99,Lee06,9Lisf,8Lipsf}
and {\it ab-initio} VMC SFs $S_{ab}$ \cite{Lap99,9Lisf,8Lipsf,wiringa}.
 } 
\begin {center}
\begin{tabular}{ p{0.8cm} p{0.8 cm} p{1.1 cm} p{1.3 cm} p{1.5 cm}p{1.4 cm}
p{0.6 cm}}
\hline
\hline
 $A$ &  $A$$-$1 & $S^{DE}$  &  M3YE & RM3YE &  $S_{exp}$ &
 $S_{ab}$ \\
 \hline
$^{3}$H    & $d$      & 1.5  &1.21 & 1.33 &    & 1.30\\
$^{3}$He   & $d$      & 1.5  &1.22 & 1.35 &    & 1.32\\
$^{4}$He   & $^{3}$He & 2.0  &1.29 & 1.42 &    & 1.50\\
$^{7}$Li   & $^{6}$He & 0.69 &0.28 & 0.33 &   0.42(4) & 0.42\\      
$^{7}$Li   & $^{6}$Li & 0.87 & 0.44& 0.46 &  0.74(11)   & 0.68\\ 
$^{8}$Li   & $^{7}$He & 1.02 &0.38 & 0.44 &   0.36(7) & 0.58\\
$^{8}$Li   & $^{7}$Li & 1.14 &0.65 & 0.77 &   & 0.97\\
$^{8}$B    & $^{7}$Be & 1.14 &0.78 & 0.91 &   0.89(7)  & 0.97\\
$^{9}$Li   & $^{8}$Li & 1.04 & 0.60 & 0.70 &   0.59(15) & 1.14\\
$^{9}$Be   & $^{8}$Li & 1.13 & 0.45  &  0.49 &   &0.73\\
$^{9}$C    & $^{8}$B  & 1.04 &  0.71 & 0.82 &   0.77(6) & 1.14\\
$^{10}$Be  & $^{9}$Li & 1.93 &  0.81 &  0.88 &  & 1.04\\
$^{10}$Be  & $^{9}$Be & 2.67 & 1.48  & 1.68  &  & 1.93\\
$^{12}$B   & $^{11}$B &  0.99 & 0.97 & 0.84 &  0.40(6) &\\
$^{12}$C   & $^{11}$B & 2.85 & 1.55 & 1.76  &  1.72(11) &\\
$^{13}$C   & $^{12}$C & 0.63 & 0.63 & 0.51 &  0.54(8)  &\\ 
$^{14}$C   & $^{13}$C & 1.87 &1.82 & 1.49 &   1.07(22) &\\
$^{14}$N   & $^{13}$N & 0.72 & 0.60 & 0.53 &  0.48(8)  &\\
$^{15}$N   & $^{14}$N & 1.48 & 1.31 & 1.06 &  0.93(15) &\\
$^{16}$O   & $^{15}$N & 2.13 & 1.57 & 1.29 &  1.27(13) &\\
\hline 
\end{tabular}
 
\end{center}
\label{table1}
\end{table}

For open shell nuclei,  $S^{IE}$  also depend on
occupancies of the single-particle 
orbits in the chosen model space, or on weights
of the $SU(3)$ and $SU(4)$ configurations in the supermultiplet shell model.
I generate these weights using  phenomenological
interaction from \cite{millener}
which gives improved spectra of $0p$ shell nuclei.
I remove the centre-of-mass motion explicitly and use $r_{osc}$ chosen as an
average of values for nuclei $A$ and $B$ derived in \cite{Jag74} from
electron scattering.
The resulting  SFs $S^{IE}$ for ground states
of the $0p$-shell nuclei, obtained with M3YE,  are compared in Table I
to $S^{DE}$ and
to SFs 
 available   from knockout 
and those transfer
reactions that use Hartree-Fock wave functions for transfer states. 
For all of them,
$S^{IE} < S^{DE}$, which clearly displays the SFs reduction phenomenon. 
However,  
$S^{IE}> S_{exp} $ for $0p_{1/2}$   and $S^{IE} < S_{exp} $ for 
$0p_{3/2}$.
Agreement between $S^{IE}$ and $S_{exp}$ 
can be improved
by tuning the $\tilde{V}^{\rm eff}$ potential. In this letter, for demonstration
purpose only, I
make the following changes to M3YE. All potentials in even partial waves
are multiplied by 1.05. This increases the SFs for d, $^3$H, $^{3,4}$He 
by 10$\%$. Then the central and spin-orbital odd components 
are multiplied by 1.7 and 
2.5 respectively,  which allows 
$S_{exp}$ for both  $^{12}$C and $^{16}$O to be reproduced. 
Increasing
odd tensor component twice    reproduces
the SF for $^{13}$C. The SFs calculated  with such a renormalised
potential, called here RM3YE, are shown in Table I.
Most SFs 
agree well with
experimental data. Detailed discussion of this
comparison will be published elsewhere.

\begin{figure}[t]
        \centerline{\psfig{figure=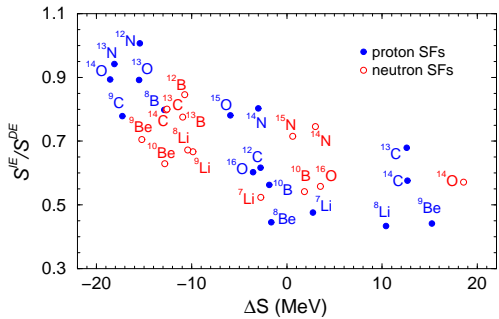,width=0.36\textwidth}
        }
\caption{
(Colour online) The ratio $S^{IE}/S^{DE}$,   calculated with RM3YE. 
}
\end{figure}

The ratio $R^{IE}_{DE}=S^{IE}/S^{DE}$, 
obtained with RM3YE,
is shown in Fig. 2 as a function of $\Delta S$.
 The decrease towards
large positive $\Delta S$ can partially be explained by   the
presence of  $\kappa$-dependent function
$G_l(r,r')$ in Eq. (\ref{ineq}). Computer calculations show that,
for fixed $\Phi_A$, $\Phi_B$ and $\tilde{ V}^{\rm eff}$,
$S^{IE}$ decreases with increasing $\kappa$. Other effects must
be also responsible for    $R^{IE}_{DE}(\Delta S)$  behaviour 
but no rigorous
explanation to it   is  yet available.

The ratio $R^{IE}_{DE}(\Delta S)$ is remarkably similar to 
  $R_s(\Delta S)$
 from \cite{Gad08}. 
This suggests that what really is measured in one nucleon
removal experiments is not $S^{DE}$ but $S^{IE}$, thus implying  that these
experiments study not  occupancies of the shell model orbits but
effective interactions $ {V}_N C_{NB}$
for occupancies fixed from other observables, such as binding energies
or nuclear
spectra. Due to the presence of the Green function in Eq. (\ref{ineq}),
$S^{IE}$ carries much more information about missing model spaces
than $S^{DE}$.
Therefore, it may be difficult to get correct values
for SFs  by overlapping  wave functions directly even if they are
obtained in a correlated {\it ab-initio}  approach. Indeed, the VMC
SFs for light nuclei are systematically larger than $S^{IE}$ calculated
in a much simpler model
with a reasonably chosen effective interaction, and, except for $^7$Li,
the VMC SFs are in a worse agreement with experiment than 
those from the present work (see Table I).


Thus,  for   fifty years SFs have been calculated in a  
procedure  of direct overlapping model wave functions
 that is sensitive only to effective interactions in truncated model space
 and does not contain
important contributions from excluded model spaces.
Calculating SFs from $I_{lj}^{IE}(r)$ generated by
Eq. (\ref{ineq}) is a more appropriate procedure that 
allows small model spaces to be used to explain the large reduction of
spectroscopic strength due to coupling to missing model spaces.
Moreover, explicitely depending on NN matrix elements both
in truncated and excluded spaces and having a guaranteed correct
asymptotic form,  $I_{lj}^{IE}(r)$ itself becomes
an interface between 
nuclear structure and nuclear reactions theories.
Incorporating Eq. (\ref{ineq}) into widely used shell model codes
and into other microscopic approaches, including {\it ab-initio} ones,
would be highly benefitial for modern nuclear physics and for 
astrophysical applications in particular.




I am grateful to D.J. Millener for providing me with his supermultiplet
shell model code and NN interactions and for helpful discussions,
M. Viviani and A. Kievsky for sending me overlap funditons of three- and
four-body systems, L. Lapik\'as for clarifying accuracy of the $^7$Li SF,
R.C. Johnson for valuable comments and help in preparing this letter.
The  support from the UK STFC grant ST/F012012/1 is thankfully
acknowledged.

\end{document}